 \documentclass[preprint2]{aastex}

\makeatletter
\def\blfootnote{\xdef\@thefnmark{}\@footnotetext}
\makeatother

\shorttitle{Evolved Nature of the B[e] Star MWC\,137}
\shortauthors{Muratore et al.}

\begin{document}

 \title{Evidence of the Evolved Nature
       of the B[e] Star MWC\,137\,$^{\star}$\blfootnote{$^{\star}$\,Based
    on observations obtained at the Gemini Observatory, which is operated by the 
    Association of Universities for Research in Astronomy, Inc., under a cooperative agreement 
    with the NSF on behalf of the Gemini partnership: the National Science Foundation 
    (United States), the National Research Council (Canada), CONICYT (Chile), the Australian 
    Research Council (Australia), Minist\'{e}rio da Ci\^{e}ncia, Tecnologia e Inova\c{c}\~{a}o 
    (Brazil) and Ministerio de Ciencia, Tecnolog\'{i}a e Innovaci\'{o}n Productiva (Argentina), 
    under program IDs GN-2011B-Q-24 and GN-2013B-Q-11.}}

\author{M.~F. Muratore\,\altaffilmark{1}}
\affil{Departamento de Espectroscop\'ia Estelar, Facultad de Ciencias Astron\'omicas y Geof\'{i}sicas, 
       Universidad Nacional de La Plata, and Instituto de Astrof\'{\i}sica de La Plata, CCT La Plata, 
       CONICET-UNLP, Paseo del Bosque S/N, B1900FWA, La Plata, Argentina}
\email{fmuratore@carina.fcaglp.unlp.edu.ar}

\author{M. Kraus and M.~E. Oksala}
\affil{Astronomick\'y \'ustav, Akademie v\v{e}d \v{C}esk\'e Republiky, Fri\v{c}ova 298, 251\,65 Ond\v{r}ejov, Czech Republic}

\author{M.~L. Arias\,\altaffilmark{2} and L. Cidale\,\altaffilmark{2}}
\affil{Departamento de Espectroscop\'ia Estelar, Facultad de Ciencias Astron\'omicas y Geof\'{i}sicas, 
       Universidad Nacional de La Plata, and Instituto de Astrof\'{\i}sica de La Plata, CCT La Plata, 
       CONICET-UNLP, Paseo del Bosque S/N, B1900FWA, La Plata, Argentina}

\author{M. Borges Fernandes}
\affil{Observat\'orio Nacional, Rua General Jos\'e Cristino 77, 20921-400 S\~ao Cristov\~ao, Rio de Janeiro, Brazil}
               
\and

\author{A. Liermann}
\affil{Leibniz-Institut f\"ur Astrophysik Potsdam (AIP), An der Sternwarte 16, 14482 Potsdam, Germany}

\altaffiltext{1}{\,Research Fellow of the Universidad Nacional de La Plata, Argentina}
\altaffiltext{2}{\,Member of the Carrera del Investigador Cient\'ifico, CONICET, Argentina}

\begin{abstract}
  The evolutionary phase of B[e] stars is difficult to establish due to the un
  certainties in their fundamental parameters. For instance, possible
  classifications for the Galactic B[e] star MWC 137 include pre-main-sequence
  and post-main-sequence phases, with a large range in luminosity.
  Our goal is to clarify the evolutionary stage of this peculiar object, and
  to study the CO molecular component of its circumstellar medium.
  To this purpose, we modeled the CO molecular bands using high-resolution $K$-band spectra.
  We find that MWC\,137 is surrounded by a detached cool ($T = 1900\pm 100$\,K)
  and dense ($N = (3\pm 1)\times 10^{21}$\,cm$^{-2}$) ring of CO gas orbiting 
  the star with a rotational velocity, projected to the line of sight, of 
  $84 \pm 2$ km$\,$s$^{-1}$.
  We also find that the molecular gas is enriched in the isotope 
  \textsuperscript{13}C, excluding the classification of the star as a Herbig Be.
  The observed isotopic abundance ratio ($ \mathrm{ \textsuperscript{12}C / \textsuperscript{13}C } = 25 \pm 2$)
  derived from our modeling is compatible with a proto-PN, main-sequence or supergiant evolutionary phase.
  However, based on some observable characteristics of MWC\,137,
  we propose that the supergiant scenario seems to be the most plausible. 
  Hence, we suggest that MWC\,137 could be in an extremely short-lived phase,
  evolving from a B[e] supergiant to a blue supergiant with a bipolar ring nebula.
\end{abstract}


\section{INTRODUCTION}
\label{intro}

 \object{MWC\,137} (V1380\,Ori) is a peculiar early-type star located in the Galactic plane
 ($l = 195.65\degr, b = -0.07\degr$), surrounded by the optical nebula Sh 2-266 (PK 195-00$\degr$1). 
 Its optical spectrum shows strong Balmer lines in emission, as well as permitted and 
 forbidden emission lines from other elements, such as 
 He\,{\sc i}, Ca\,{\sc ii}, [O\,{\sc i}], [O\,{\sc ii}], [N\,{\sc ii}], and [S\,{\sc ii}]
 \citep{hamman1992b,zickgraf2003,hernandez2004}. 
 This emission-line spectrum, together with the presence of a large infrared excess 
 \citep{frogel1972,cohen1975,hillenbrand1992}
 usually attributed to free-free emission and thermal emission from dust, is responsible 
 for the classification of this object as a B[e] star 
 \citep{allen1973,bergner1995,esteban1998,marston2008}.
  
 Both the nature and evolutionary status of this object remain uncertain.
 Based on optical spectra and infrared photometry, \citet{frogel1972} conclude
 that MWC\,137 is probably an early-type emission-line star surrounded by nebulosity. 
 \citet{cohen1975} arrive to the same conclusion from the analysis of optical, 
 infrared and radio data.
 \citet{sabbadin1981A&A....94...25S} present photographic and spectroscopic observations, 
 and propose that Sh 2-266 is likely a ``low excitation nebula ejected and excited by a 
 peculiar star".  
 Due to the appearance of its emission-line spectrum, this star has been included in many
  catalogs and articles on Herbig Ae/Be stars \citep{hillenbrand1992,berrilli1992,the1994}.
 Based on photometric and spectroscopic studies, \citet{miroshnichenko1994} and 
 \citet{bergner1995} conclude that this object is an early-B type star near the zero-age 
 main sequence (ZAMS), although the latter state that it is not clear whether it is a 
 pre-main-sequence object, or a more massive star that has just left the main sequence.  
 Some authors, however, argue against a pre-main-sequence nature of the object based 
 on its high luminosity,
 the shape of the H$\mathrm{\alpha}$, Na\,{\sc i} D and He\,{\sc i} 5876 \AA~line profiles, 
 and the fact that the radio emission at 10 GHz is much stronger than in any other Herbig Be 
 star \citep{herbig1972,finkenzeller1984}.
 On the other hand, \citet{manchado1989} classify the star as a late B-type supergiant based 
 on a photometric analysis, while \citet{esteban1998}, using high-resolution spectroscopy 
 and H$\mathrm{\alpha}$ imaging, conclude that the star is an evolved early B supergiant. 
 While most studies agree on assigning an early spectral type to the star, ranging from
 late O \citep{brand1994,rudolph1996} to early B \citep{hillenbrand1995,miroshnichenko1994},
 the determination of the luminosity class is still subject to debate.
 This is due to the fact that few photospheric absorption features are seen in the stellar spectrum. 
 In addition, the amount of circumstellar extinction is not known,
 complicating the distance and luminosity determination.
 
 The nebula around MWC\,137 was first classified as an H\,{\sc ii} region by 
 \citet{sharpless1959}, who designated it as Sh 2-266.
 Values for the kinematical distance, obtained under the assumption that the H\,{\sc ii} region
 was associated with a molecular cloud with measured CO velocity, include: 
 12.6 $\pm$ 3.2 kpc \citep{fich1984,wouterloot1988}, 9.6 $\pm $ 4.8 kpc \citep{fich1991}, 
 and 10.98 kpc \citep{brand1994}.
 This object was also cataloged as the planetary nebula (PN) PK 195-00$\degr$1 by \citet{perek1967}, 
 and although subsequent work questioned this classification \citep{frogel1972,cohen1975}, 
 it continued to appear in many PNe studies. 
 Distances derived with methods developed specifically for PNe range from 870 pc 
 \citep{amnuel1984} to 1.62 kpc \citep{zhang1995}.
 Herbig Be studies usually adopt distances of the order of 1 kpc, leading to luminosities
 in the range $\log \, (L/L_{\sun}) \sim 4.2 - 4.5$ \citep[e.g.][]{hillenbrand1992}.
 \citet{esteban1998} use high-resolution spectroscopy to measure the radial velocity 
 of different nebular lines, and derive a kinematical distance $d \ge 6$ kpc. With this 
 lower limit for the distance, the authors obtain a luminosity $\log (L/L_{\sun}) \ge 5.4$, 
 compatible with a supergiant evolutionary stage.

 A way to disentangle the nature of MWC\,137 (and of unclassified B[e] stars in general) 
 is to analize the properties of the circumstellar material (spatial distribution, 
 composition, physical conditions, and kinematics), since they are directly related 
 to the evolutionary stage of the underlying object.
 The first overtone CO band emission visible in the $K$ band constitutes an extremely 
 valuable tool for obtaining information about the star and the surrounding material. 
 This feature has been observed in a variety of objects, including 
 pre-main-sequence \citep[e.g.][]{carr1995,najita1996} and evolved stars, 
 both low-mass \citep[e.g.][]{gorlova2006,gledhill2011} and 
 high-mass \citep[e.g.][]{mcgregor1988a,mcgregor1988b,morris1996,oksala2013}.
 It originates in the inner edge of a molecular circumstellar disk, 
 making it an ideal indicator for both the disk parameters (temperature and density) 
 and kinematics.
 In addition, the chemical composition of the molecular material around evolved stars
 can provide information about the evolutionary stage of the star at the time of mass ejection.
  
 During the evolution of stars, heavy elements produced in the stellar interior are 
 transported to the surface via rotation and mixing processes.
 Since the stellar surface abundance of \textsuperscript{13}C increases with the
 age of the star \citep{kraus2009}, this element is a very valuable tracer for stellar evolution. 
 Mass loss deposits the chemically enriched matter into the circumstellar environment, 
 where it condenses into molecules.
 The detection of \textsuperscript{13}CO emission bands with high-resolution spectrographs
 reflects the existence of a substantial amount of \textsuperscript{13}C in the 
 circumstellar medium \citep{kraus2009}. 
 The idea of using \textsuperscript{13}CO as an indicator of the stellar evolutionary phase
 was proposed by \citet{kraus2009}, and applied by \citet{liermann2010}, \citet{muratore2010}, 
 \citet{kraus2013}, and \citet{oksala2013}, to confirm the evolved nature of several objects, 
 including B[e] supergiants and yellow hypergiants.
 This method is extremely useful for stars with ambiguous classification, such as MWC\,137, 
 because it allows to distinguish evolved stars from young, pre-main-sequence objects.

 In this work we investigate the CO emission from MWC\,137 
 using high-resolution near-infrared spectroscopy.
 The presence of \textsuperscript{12}CO band emission was reported by 
 \citet{muratore2012} and \citet{oksala2013}. 
 In addition, \citet{oksala2013} detected \textsuperscript{13}CO emission bands,
 which were confirmed by \citet{liermann2014MNRAS.443..947L}.
 Based on the amount of \textsuperscript{13}CO present in the $K$-band spectrum,
 these authors suggest that the star is not a pre-main-sequence object.
 Given that their result was based on low-resolution data, it is important
 to confirm the \textsuperscript{13}C enrichment,
 since it could constrain the evolutionary state of the star.  
 In addition, it could also provide evidence about the origin of the
 CO disk and gaseous nebula (accreted vs ejected material).
 When the disk is formed of material ejected by a
 hot massive star, the presence of molecules and dust 
 within a few tens of AU from the star is not expected due to the high UV radiation flux.
 Therefore, it is crucial to investigate the physical and kinematical properties
 of such circumstellar envelopes, since
  the formation and survival of dust and molecules under harsh conditions is one of the
 unresolved issues of stellar evolution.
 Furthermore, it is important to know 
 the amount of mass ejected by the star in previous stages of the evolution, since it could affect the properties
 of the object in the following phases (e.g., evolution time-scale, fate of the star).

\section{OBSERVATIONS}
\label{observations}
 
  High resolution ($R \sim 18000$) $K$-band spectroscopic observations of MWC\,137 were obtained
  in 2011 and 2014
  using the Gemini Near-Infrared Spectrograph (GNIRS) mounted on the Gemini North telescope. 
  The observations taken in 2011, which included the first two \textsuperscript{12}CO band heads,
  were aimed at studying the kinematics and physical properties of the 
  gas. 
  The goal of the second set of observations, obtained in 2014, was to confirm the presence of 
  \textsuperscript{13}CO emission and model it in detail.  
  The spectral range covered by the 2011 GNIRS observations was also included to
  check the possible variability of the \textsuperscript{12}CO bands. 
  Details related to the observing programs can be found in Table \ref{table:obs}.
%
 \begin{table*}
 \begin{center} 
 \caption{Details on  MWC 137  $K$-band observations.\label{table:obs}} 
 \begin{tabular}{c c c c c}
 \tableline
  Program ID   & Instrument  &    Date    & Spectral Range & CO Coverage \\
               &             &            &   ($\mu$m)     &             \\
 \tableline
 GN-2011B-Q-24 &    GNIRS    & 2011-10-07 & $2.28 - 2.34$  & first two \textsuperscript{12}CO band heads \\ 
 GN-2013B-Q-11 &    GNIRS    & 2014-03-21 & $2.28 - 2.39$  & four \textsuperscript{12}CO band heads,     \\
               &             &            &                & first two \textsuperscript{13}CO band heads \\
 \tableline
 \end{tabular}
\end{center}
\end{table*}
%
 The observations were performed in longslit mode, using the long camera, the 110 l/mm grating,
 and the 0.10 arcsec slit. 
 A central wavelength of $2.312$ $\mu$m was chosen in 2011, while   
 two wavelength intervals, centered at $2.312$ and $2.36$ $\mu$m, were 
 observed in 2014. 
 Observations were taken with an ABBA nod pattern along the slit in order to
 remove sky emission. All the steps of the reduction process were made using
 IRAF\footnote{IRAF is distributed by the National Optical Astronomy Observatories,
    which are operated by the Association of Universities for Research
    in Astronomy, Inc., under cooperative agreement with the National
    Science Foundation.}
 software package tasks. Reduction steps include AB pairs subtraction, flat field
 correction, telluric correction, and wavelength calibration. The telluric correction
 was performed using a B-type telluric standard star observed at similar airmass.
 After applying the corrections for heliocentric and systemic velocities, the 
 continuum was used to normalize the data, and finally, it was subtracted to obtain
 pure emission spectra.
  
 \begin{figure*}
   \centering
   \includegraphics[width=9cm,angle=270]{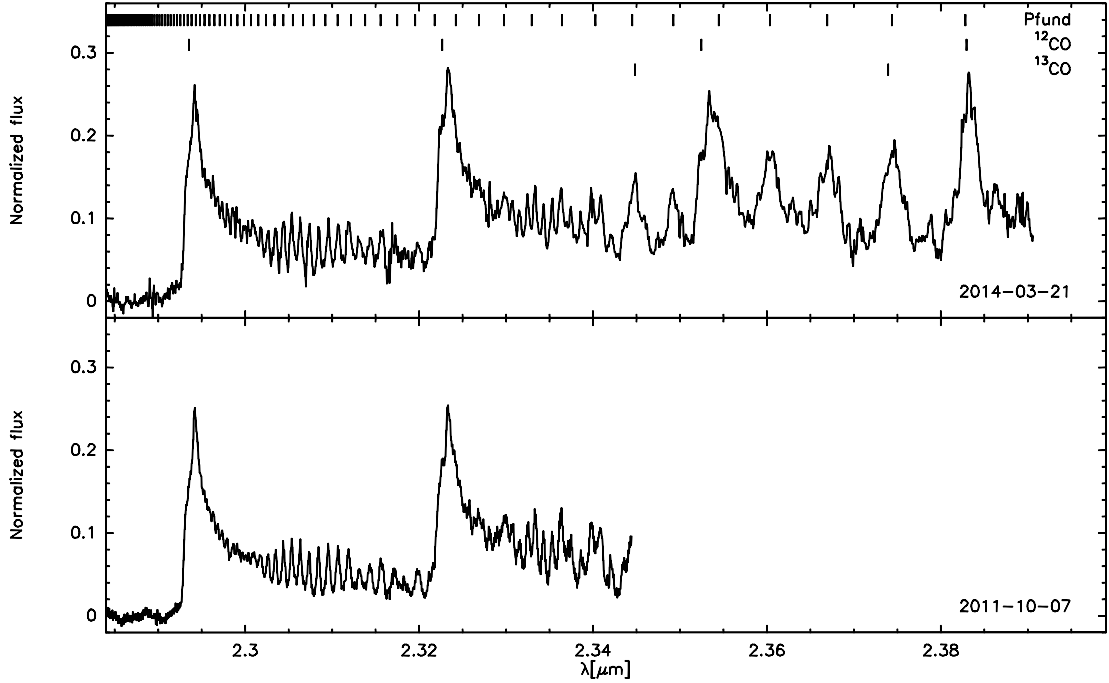} 
   \caption{Normalized, continuum-subtracted GNIRS spectra from 2011
            (bottom panel) and 2014 (top panel). The position of the
             \textsuperscript{12}CO and \textsuperscript{13}CO band heads,
            and the wavelengths corresponding to the Pfund transitions 
            are indicated.\label{Fig-obs}}
 \end{figure*}
 
 Fig.~\ref{Fig-obs} shows the normalized, continuum-subtracted GNIRS spectra
 from 2011 (bottom panel) and 2014 (top panel), which clearly display the CO first
 overtone bands in emission. 
 In addition, the spectrum from 2014 shows emission lines from the hydrogen Pfund
 series superimposed on the molecular bands. These lines are also present in the 
 2011 spectrum, but they are not so easily distinguished in this spectral range. 
 The ticks in the figure mark the position of the \textsuperscript{12}CO and
 \textsuperscript{13}CO band heads, and the wavelengths corresponding to the 
 Pfund transitions.
 Both spectra look similar, which implies that the CO emission has 
 not changed significantly in the last two years.
\section{SPECTRAL MODELING}

  The physical conditions and kinematics of the CO emitting region 
 can be determined by modeling the emission bands present in our $K$-band
 spectra of MWC\,137.
  However, the amount of \textsuperscript{13}CO present in the circumstellar
 environment can only be determined by modeling the spectrum obtained in 2014. 
  Given that the CO emission did not change significantly between the two epochs, 
 in this section we describe in detail the modeling and the results obtained from the 
 2014 spectrum,
 since its spectral coverage allows to study both, the 
 \textsuperscript{12}CO and the \textsuperscript{13}CO band emission.
 The contribution of the Pfund lines to the observed spectrum must also be taken
 into account to properly fit the CO band emission.
 At the end of this section we compare the results obtained
 from fitting the spectra taken in both epochs
 to check for possible differences.

 \subsection{Description of the Models}
 
 The Pfund emission lines do not appear to be double-peaked, hence they do not originate
 in a rotating disk, but most probably in a high-density ionized wind or shell. 
 We use the code developed by \citet{kraus2000}, which, following Menzel's recombination 
 theory \citep{menzel1938}, computes the emission from these lines under the assumption 
 that they are optically thin.    
 The line profiles are assumed to have a Gaussian shape, characterized by the velocity 
 $v_{\, \rm gauss,\rm Pf}$. 
 The physical properties of the gas are given by the electron density $n_e$, the proton 
 density $n_p$, which for simplicity we set equal to $n_e$, and the electron temperature $T_e$. 
 Since the population of the higher levels can be prevented by pressure ionization effects
 in high-density media, the code allows to specify the quantum number $\rm n_{max}$ 
 of the maximum transition included in the computation.
 This will be used to determine a lower limit for $n_e = n_p$. 
 
\begin{figure}
    \plotone{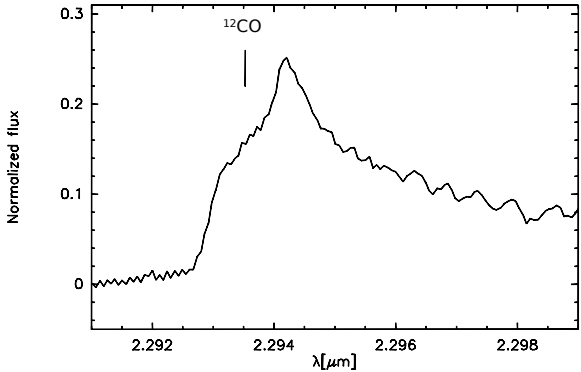}
    \caption{Normalized, continuum-subtracted GNIRS spectrum showing in detail the rotationally 
             broadened structure of the first \textsuperscript{12}CO band head. The tick marks 
             the position of the band head.}
    \label{Fig-gnirs-bandhead}
 \end{figure}

 The shape of the first \textsuperscript{12}CO band head gives an indication of the 
 kinematical broadening of the individual CO lines.
 Fig.~\ref{Fig-gnirs-bandhead} shows a blowup of the GNIRS spectrum in the region
 of the first \textsuperscript{12}CO band head. 
 The blue shoulder and the red peak are both characteristic of emission originated 
 in a rotating ring of material \citep[e.g.][]{carr1993,carr1995,najita1996,kraus2000}.
%
%
 We model the CO bands using the \textsuperscript{12}CO disk code of \citet{kraus2000},
 modified to include \textsuperscript{13}CO \citep{kraus2009,oksala2013}.
 This code computes the \textsuperscript{12}CO and \textsuperscript{13}CO emission 
 arising from a narrow ring of material with constant temperature $T_{\rm CO}$, 
 column density $N_{\rm CO}$, rotational velocity projected to the line of sight 
 $v_{\rm rot, CO} \sin i$, and a given isotopic abundance ratio 
 $ \mathrm{ \textsuperscript{12}C / \textsuperscript{13}C }$.
 These choices reflect the fact that the CO emission is usually found to arise 
 from a disk, and that the hottest component, assumed to be located in a relatively 
 narrow region, is mainly responsible for the observed CO spectrum
 \citep{kraus2000,kraus2009,liermann2010,cidale2012,kraus2013}.
 
 To derive the characteristics of the CO emitting region, we model simultaneously 
 the contribution to the $K$-band spectrum of both the CO band emission and the Pfund line emission.

\subsection{Results}
\label{results}

 The width and shape of the Pfund lines are well fit by a Gaussian profile 
 with $v_{\, \rm gauss,\rm Pf} = 170 \pm 10$ km$\,$s$^{-1}$.
 Since the separation between the individual lines decreases as the transitions 
 become higher, the Pfund lines start to blend towards shorter wavelengths and 
 form a quasi-continuum. 
 The wavelength where this occurs depends on the width of the lines and the 
 resolution of the spectrum. 
 Taking into account the broad profiles of the Pfund lines and the resolution of 
 the GNIRS spectrum, this quasi-continuum should start at $\lambda < 2.32 \, \mu$m, 
 and extend down to the Pfund discontinuity at $\lambda \sim 2.28 \, \mu$m. 
 However, we find from the modeling that the contribution of the Pfund series is 
 limited to a certain number of transitions, being Pf(55) the maximum transition present.  
 The fact that the cut off occurs at such a low transition indicates a very high 
 electron density. 
 The lowest value of the density compatible with $\rm n_{max} = 55$ can be estimated
 by considering pressure ionization effects, which gives $ n_e > 10^{12} $ cm$^{-3}$ 
 \citep{kraus2000PhD}. 
 The Pfund lines visible in the GNIRS spectrum correspond to transitions associated 
 with quantum numbers in the range 24 - 55. 
 Given the high electron density, the deviation from LTE is negligible for these members
 of the series \citep[see][]{kraus2000PhD}. 
 Furthermore, the relative intensities of these lines are not sensitive to the electron
 temperature, so we modeled the final Pfund spectrum with $ n_e = 10^{13} $ cm$^{-3}$ 
 and a fixed electron temperature $T_e = 10\,000$ K, representative of an ionized nebula.
 
 The kinematics of the CO gas can be easily extracted from the shape of the first band head.      
 Fig.~\ref{Fig-gnirs-bandhead} shows the rotationally broadened structure, and modeling 
 yields a rotational velocity, projected to the line of sight, 
 of $v_{\rm rot, CO} \sin i = 84 \pm 2$ km$\,$s$^{-1}$.  
 The structure of the \textsuperscript{12}CO bands is sensitive to the temperature
 and column density of the gas in the emitting region, as described in \citet{kraus2000PhD,kraus2009}.
 Due to the high resolution of the GNIRS spectrum, the temperature can be determined
 from the appearance of the features in the region in front of the second band head. 
 This results in a value of $T_{\rm CO} = 1900 \pm 100$ K. 
 Determination of the column density requires a wider spectral range than the one 
 used to derive the temperature. 
 From the modeling of the four CO band heads visible in the last spectrum
 taken with GNIRS we obtain $N_{\rm  CO} = (3 \pm 1) \times 10^{21} $ cm$^{-2}$. 
 Considering that the dissociation temperature for the CO molecule is $\sim$ 5000 K, 
 the derived temperature implies that the material producing the band head structure 
 is rather cool, suggesting it is detached from the star.

\begin{figure*}
   \centering
   \includegraphics[width=11cm,angle=270,clip=true]{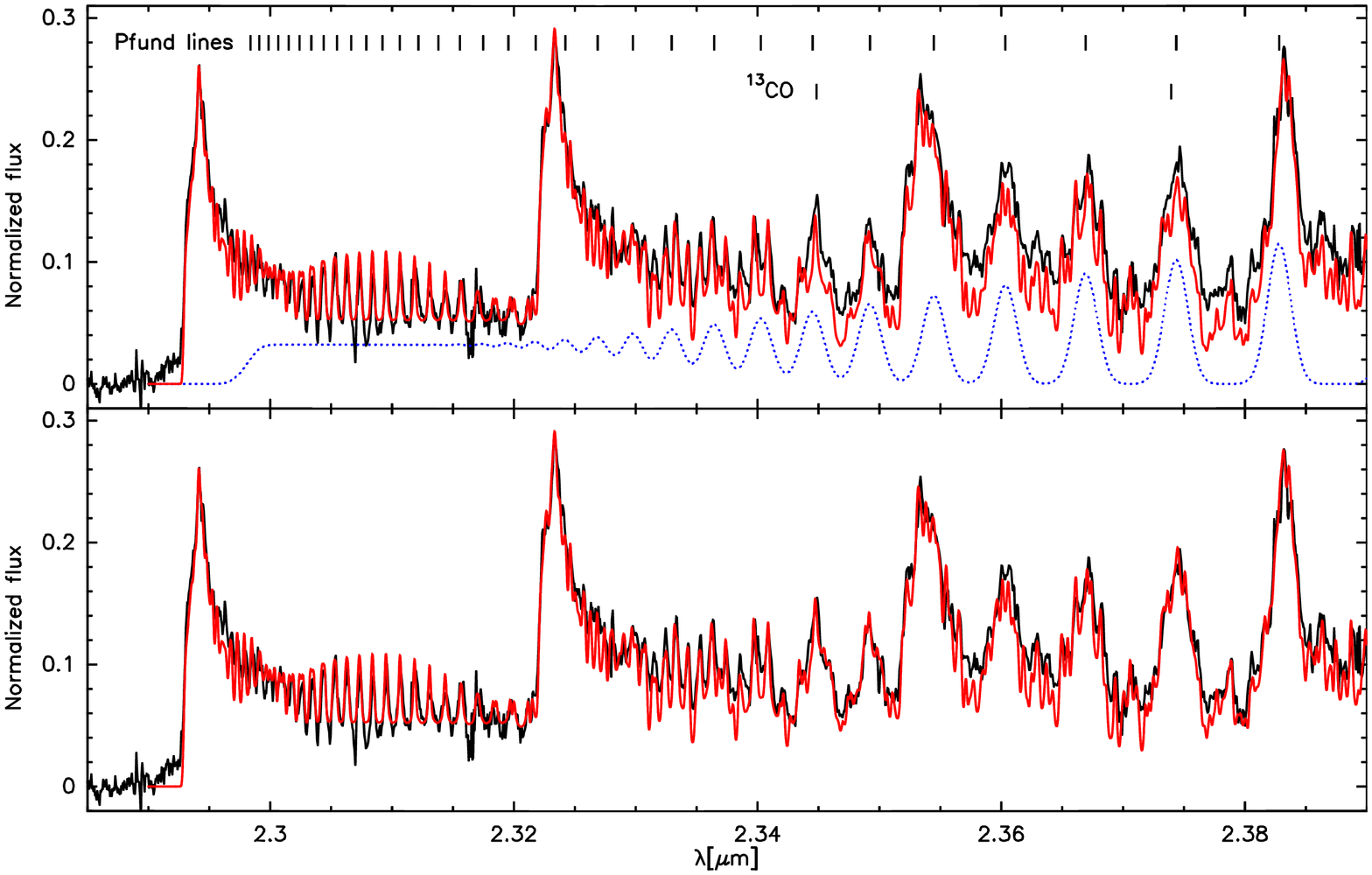}
   \includegraphics[width=6.3cm,angle=270,clip=true]{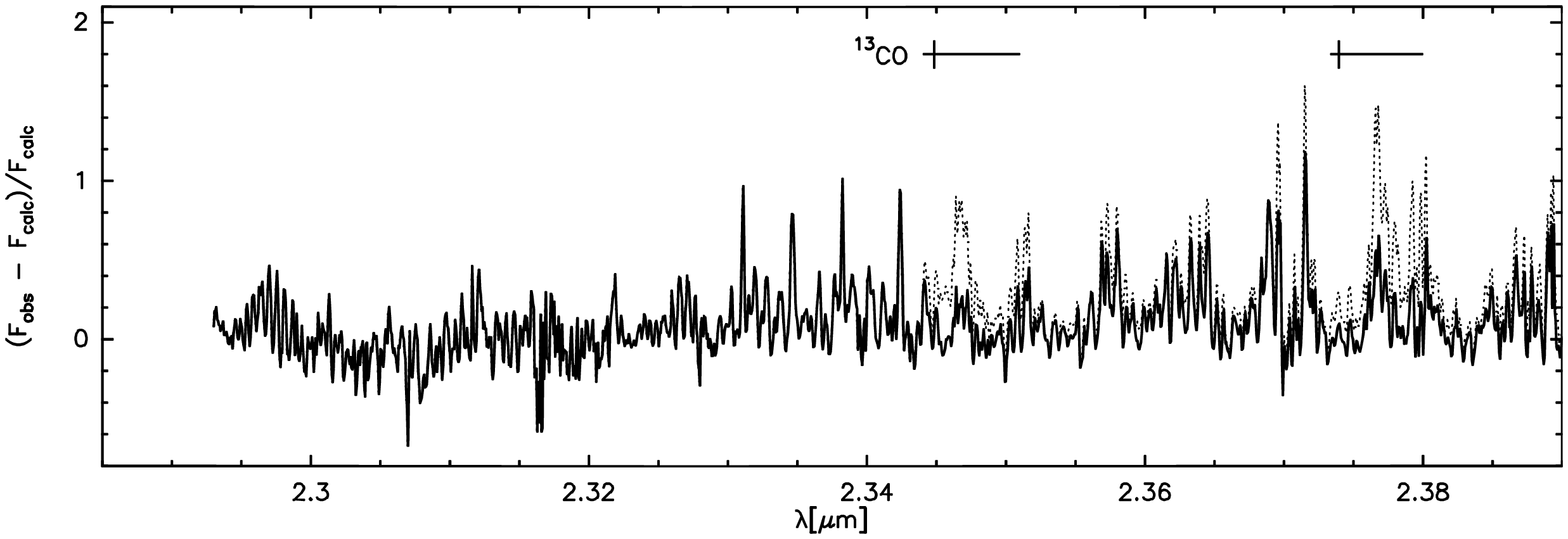}
   \caption{\textit{Top and middle panels:} Normalized, continuum-subtracted GNIRS spectrum (black)
            along with the best-fit models (red) obtained for the combined Pfund\,+\,CO emission
            assuming a ratio of $ \mathrm{ \textsuperscript{12}C / \textsuperscript{13}C } = 90$ (top)
            and $ \mathrm{ \textsuperscript{12}C / \textsuperscript{13}C } = 25$ (middle).  
            The synthetic Pfund emission spectrum (blue dotted line) included in the models 
            is shown separately in the top panel. 
            The position of the Pfund transitions and the \textsuperscript{13}CO band heads is 
            also indicated.
            \textit{Bottom panel:} Residuals of the fits for the models with $ \mathrm{ \textsuperscript{12}C / \textsuperscript{13}C } = 90$
            (dotted black line) and $ \mathrm{ \textsuperscript{12}C / \textsuperscript{13}C } = 25$ (black solid line).
            The ``crosses'' mark the wavelength and FWHM of the two \textsuperscript{13}CO band head structures. \label{Fig-fit}}  
 \end{figure*}

 With fixed values for temperature, column density, and rotational velocity, 
 we first compute the CO emission spectrum with a carbon isotopic abundance 
 ratio $ \mathrm{ \textsuperscript{12}C / \textsuperscript{13}C } = 90$.  
 According to stellar evolution models, this value 
 corresponds approximately to the beginning of the main sequence
 \citep[see, e.g., ][]{ekstrom2012}. 
 The top panel of Fig.~\ref{Fig-fit} shows the resulting Pfund + CO model 
 (red solid line) superimposed on the observed emission spectrum (black solid line).
 The pure Pfund emission model is also included in this plot (blue dotted line).
 The ticks mark the wavelengths corresponding to the Pfund transitions and the 
 position of the \textsuperscript{13}CO band heads.
 %
%
%
 A closer inspection of the fit shows that some flux appears to be missing
 redward of $\lambda \sim 2.34 \, \mu$m. 
 Models with lower and higher column densities are not able to reproduce 
 the continuum level and/or the intensity of the band heads. 
 Thus, the only way to improve the fit is to increase the contribution 
 of \textsuperscript{13}CO, or in other words, lower the isotopic abundance ratio. 
 The best fit is obtained for a value 
 $ \mathrm{ \textsuperscript{12}C / \textsuperscript{13}C } = 25 \pm 2$. 
 The middle panel of Fig.~\ref{Fig-fit} shows this model (red solid line) together
 with the GNIRS spectrum (black solid line).
 Comparison between both fits clearly shows that the isotope \textsuperscript{13}CO 
 contributes to the $K$-band emission.
 To help visualize the difference between the models in the regions of the \textsuperscript{13}CO
 band heads, a plot showing the residuals of the fits was included in the bottom panel.
 The dotted curve corresponds to the residual of the fit for the model with $ \mathrm{ \textsuperscript{12}C / \textsuperscript{13}C } = 90$
 (i.e., without \textsuperscript{13}CO), and the solid line represents the residual of the fit for the model with
 $ \mathrm{ \textsuperscript{12}C / \textsuperscript{13}C } = 25$ (i.e., with \textsuperscript{13}CO).
 The ``crosses'' mark the wavelength and FWHM of the two \textsuperscript{13}CO band head structures.
 The complete set of derived Pfund and CO parameters is listed in \mbox{Table \ref{table:param}}.

 The variation of the \textsuperscript{12}C and \textsuperscript{13}C surface abundances 
 predicted by the stellar evolution models of \citet{ekstrom2012} implies that the surface
 ratio $ \mathrm{ \textsuperscript{12}C / \textsuperscript{13}C }$ drops from an initial 
 value of $\sim 90$ to less than $\sim 20-30$ by the end of the main sequence, depending 
 on the initial mass of the star.
 The value of $25 \pm 2$ obtained for MWC\,137 confirms earlier suggestions by \citet{oksala2013},
 who obtained a tentative value of $25 \pm 5$ based on a low-resolution spectrum. 
 This indicates that the star is definitively not a pre-main-sequence object,
 validation that the Herbig Be classification can finally be discarded.
 This important result still holds when other stellar evolution models are considered,
 since the initial value and the variation of the carbon isotopic ratio 
 along stellar evolution are similar in all of them  
 \citep[see, e.g., the tracks from][]{brott2011A&A...530A.115B,chieffi2013ApJ...764...21C}.
 
\begin{table}
\begin{center} 
\caption{Best fit model parameters.\label{table:param}}
\begin{tabular}{l l}
\tableline
$v_{\rm gauss,\rm Pf}$ & $170 \pm 10$ km$\,$s$^{-1}$ \\
$\rm n_{max}$ & 55 \\
$ n_e $ & $ 10^{13} $ cm$^{-3}$ \\
$T_e$ & 10\,000 K \\
\tableline
$v_{\rm rot, CO} \sin i$ & $84 \pm 2$ km$\,$s$^{-1}$ \\
$T_{\rm CO}$ & $1900 \pm 100$ K \\
$N_{\rm  CO}$ & ($3 \pm 1) \times 10^{21} $ cm$^{-2}$ \\
$ \mathrm{ \textsuperscript{12}C / \textsuperscript{13}C }$ & $25 \pm 2$ \\
\tableline
\end{tabular}
\end{center}
\end{table}

\begin{figure*}
   \centering
   \includegraphics[width=15cm,angle=0]{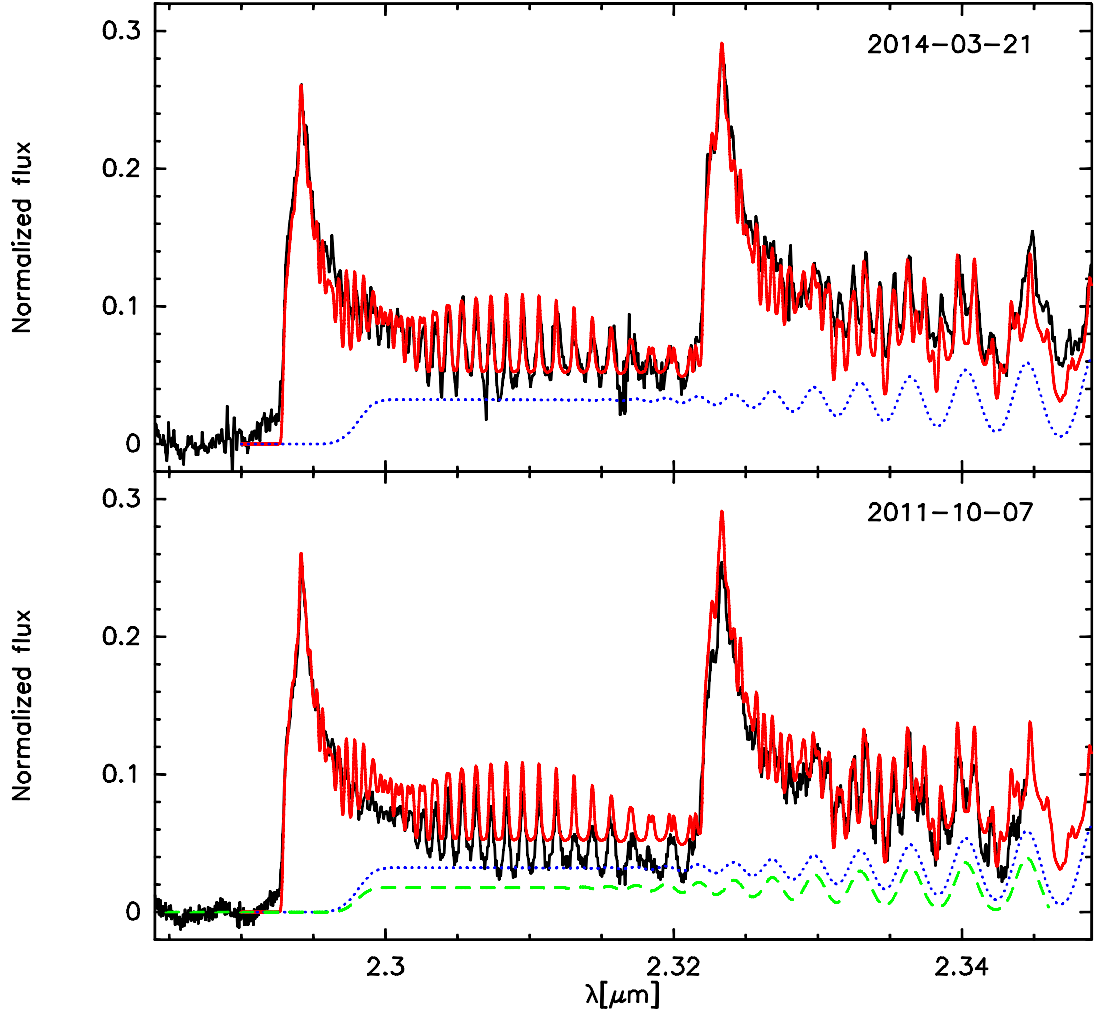}
   \caption{Normalized, continuum-subtracted GNIRS spectrum from 2014 (top panel, black solid line)
            along with the best-fit CO + Pfund model (red solid line).
            The contribution of the Pfund emission is shown in blue dotted line.
            The best-fit to the 2014 spectrum is shown in the bottom panel superimposed on the 2011 spectrum.
            The contribution of the Pfund emission to the 2011 and 2014 spectra are shown in green dashed and blue dotted lines, respectively.          \label{Fig-Pfund}}
\end{figure*}

 It is worth mentioning that the CO parameters obtained from the fits
 to the 2011 and 2014 spectra are the same. However, an increase
 in the contribution of the Pfund emission between these two epochs was found.
 The top panel of Fig.~\ref{Fig-Pfund} shows the best-fit CO\,+\,Pfund
 model (red solid line) superimposed on the spectra (black solid line) taken in
 2014. The same fit is overplotted in the bottom panel to the spectrum taken
 in 2011. Obviously, the shape of the observed spectrum is the same as in 2014, 
 but the fit redward of $\sim 2.3\,\mu$m overestimates the global intensity.
 To reduce the flux in this region, we found that a reduction in the
 contribution of the Pfund emission (dashed line versus dotted line) is 
 necessary, while the CO model remains unaltered. The difference in these 
 Pfund contributions might indicate a slight increase in the wind emission
 between the two epochs, although the cause of this variability is not so clear.
 On the other hand, the shape and intensity of the CO emission bands do not appear
 to have changed, which might point to a certain stability of the disk or ring structure
 where they originate. This kind of stability is usually associated with Keplerian rotation,
 but more observations are needed to confirm the 
 non-variable character of the CO emission.

%
%
\section{DISCUSSION}

  From the \textsuperscript{13}C enrichment of the circumstellar material detected 
 in the $K$-band spectrum, it is clear that MWC\,137 cannot be a pre-main-sequence object.
 Considering the other classifications reported by different authors and
 the variation of the $ \mathrm{ \textsuperscript{12}C / \textsuperscript{13}C }$
 ratio along the stellar evolution, in this section we analyze 
 which evolutionary scenarios are 
 compatible with the value of the isotopic abundance ratio derived from our modeling.
  After giving a quick overview of some of the 
  differences and similarities between MWC\,137 
  and objects in each of these evolutionary phases,
  we briefly discuss
  whether it is possible to constrain 
  the evolutionary stage of this star
  even further
  based on the results of our data.

 Although the classifications found in the literature 
 are based on assumptions of different nature,
 their plausibility 
 depends strongly on the distance to the object, which is still uncertain.
 Unfortunately, our data cannot support neither confirm the distances previously estimated. Even though some distance estimates found in the literature might be more reliable than others, such a discussion is out of the scope of this work. Thus, our discussion will only be based on the value of the carbon isotopic ratio.

 \subsection{$\mathrm{\textsuperscript{12}C/\textsuperscript{13}C}$ Ratio and Stellar Evolution}

\begin{figure*}
     \includegraphics[width=16.5cm]{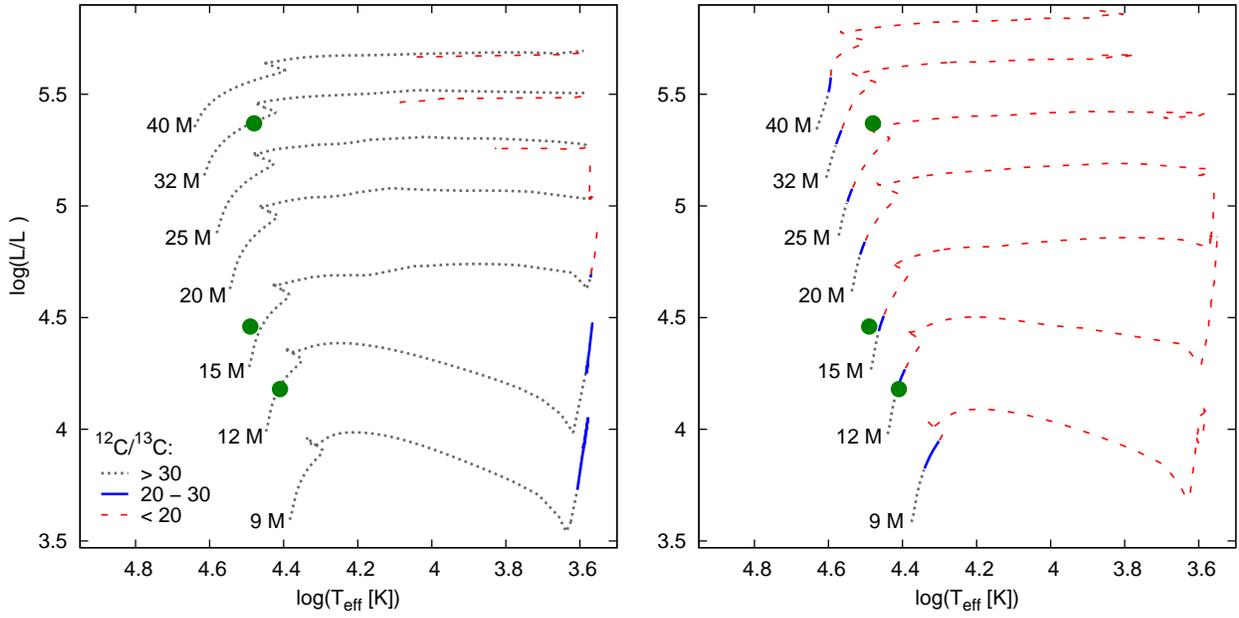} 
      \caption{HR diagrams showing the location of MWC\,137 (green spots)
       according to the stellar parameters listed in Table \ref{table:HR}.
       The evolutionary tracks from \citet{ekstrom2012} for non-rotating (left panel)
       and rotating (right panel) stars of solar metallicity are also shown. 
       The rotating tracks were computed for an initial rotation rate
       $v_{\rm ini}/v_{\rm crit} = 0.40$. The change of the 
       $ \mathrm{ \textsuperscript{12}C / \textsuperscript{13}C }$ ratio during
       stellar evolution is indicated (dashed line $\mathrm{ \textsuperscript{12}C / \textsuperscript{13}C } < 20$, solid line $ 20 < \mathrm{ \textsuperscript{12}C / \textsuperscript{13}C } < 30$, dotted line $\mathrm{ \textsuperscript{12}C / \textsuperscript{13}C } > 30$).\label{Fig-HR}}
    \end{figure*}

 Table \ref{table:HR} lists effective temperature and luminosity determinations 
 taken from the literature. 
 The classification and distance assumed by the different authors 
 are also included.
 Fig.~\ref{Fig-HR} shows the location of the star in the HR diagram based on 
 these determinations together with the evolutionary tracks from \citet{ekstrom2012} for non-rotating (left panel) and rotating (right panel) stars of solar metallicity. 
 The rotating evolutionary tracks were computed for an initial rotation rate 
 $v_{\rm ini}/v_{\rm crit} = 0.40$.
   The accuracy of these well-known stellar evolution models 
   is enough for our purpose, since the uncertainty in the stellar parameters 
   is by far larger than the difference between evolutionary tracks from different authors 
   \citep[a comparison between stellar evolution models can be found in][]{martins2013A&A...560A..16M}. 
 From the tables provided by \citet{ekstrom2012} we compute the 
 $ \mathrm{ \textsuperscript{12}C / \textsuperscript{13}C }$ ratio along the stellar
 evolution, and indicate in the figures those parts of the tracks (solid blue line) that are compatible
 with the ratio of $25 \pm 2$ derived for MWC\,137.
  
 Taking into account the classifications listed in Table \ref{table:HR}, all the 
 positions of the star in the HR diagram correspond to ratios higher than 30 for 
 the non-rotating case, as can be seen in the left panel of Fig.~\ref{Fig-HR}. 
 This is clearly inconsistent with the value of $25 \pm 2$ obtained from our fit.
 The right panel of this figure shows that the lower luminosity positions are compatible 
 with  a main-sequence star of initial mass between 12 and 16 M$_{\sun}$, and initial 
 rotation rate similar to the value used to compute the stellar evolution models.
 The higher luminosity position could correspond to a supergiant star of initial mass 
 of at least 25 - 32 M$_{\sun}$, but with a lower initial rotation rate than the one 
 that was used to compute these tracks.  
 The possibility of the star being in the PN phase cannot be analyzed in the same way 
 because there are no determinations of effective temperature or luminosity made under 
 that assumption. However, if we consider the position in the HR diagram corresponding 
 to the lowest luminosity, we find that it could be compatible with a star of initial 
 mass around 9 M$_{\sun}$, with a very low initial rotation rate, that expelled the 
 enriched material in a previous phase (AGB and post-AGB) and is now evolving towards 
 the PN phase. 
 In the following section we analyze each of these possibilities 
 and their implications. 

 \begin{table*}[h!]
 \begin{center} 
 \caption{Stellar parameters for MWC\,137 taken from the literature.\label{table:HR}} 
 \begin{tabular}{c c c c c}
 \tableline
 $\log \rm T_{\rm eff}$ &    $\log \rm L$   & Ref. & Classification\,\tablenotemark{a} &  d \\
          (K)           &  ($\rm L_{\sun}$) &      &        & (kpc)  \\
 \tableline
 4.41 & 4.18 & 1 & PMS / MS &  1.3 \\ 
 4.49 & 4.46$\pm$0.04\,\tablenotemark{b} & 2 & PMS / MS & 1.3 \\
 4.48 & 5.37\,\tablenotemark{c} & 3 & SG & 6\,\tablenotemark{c} \\
 \tableline
 \end{tabular}
 \tablenotetext{a}{\,PMS = pre-main-sequence; MS = main-sequence; SG = supergiant.}
 \tablenotetext{b}{\,Values in the references range from 4.42 to 4.5.}
 \tablenotetext{c}{\,Lower limit according to \citet{esteban1998}.}
  \tablerefs{
 (1) \citet{alonso-albi2009};     
 (2) \citet{hillenbrand1992,hillenbrand1995,testi1998,verhoeff2012};     
 (3) \citet{esteban1998}.}     
\end{center}
\end{table*}

\subsection{Possible Scenarios}

\subsubsection{Planetary Nebula}

 The luminosity corresponding to the lowest position in the HR diagram, 
 $\log \, (L/L_{\sun}) = 4.18$, is relatively high for PNe but still 
 compatible with this class of objects 
 \citep[see evolutionary tracks from, e.g.,][]{paczynski1970,paczynski1971,vassiliadis1994,bloecker1995}.
 Contrary to what is expected for typical PNe, the [O\,{\sc iii}] emission 
 in MWC\,137 is very weak \citep{frogel1972,sabbadin1981A&A....94...25S}. 
 However, \citet{kraus2005} found that the [O\,{\sc iii}] emission from the PN Hen 2-90 
 is not as strong as expected for normal PNe, and explain this with a very low O abundance.
 Another possibility is that the temperature of the star 
 is not yet high enough to ionize O\,{\sc ii}. 
 Central stars of PNe with such low temperatures might be at the (early) beginning 
 of the PN phase, so in other words, they might be young PNe. 
 However, these stars are often considered to be ``transition objects'' 
 evolving from the AGB to the PN phase, and they are referred to as pre- or proto-PNe.
 One of the best studied proto-PNe, M 1-92 (Minkowski's Footprint) was also reported 
 to show very weak [O\,{\sc iii}] emission \citep{bujarrabal1998A&A...331..361B}.
 Furthermore, the sizes of the nebulae around this object and MWC\,137 are comparable, 
 with a total extension of 0.2 pc \citep{bujarrabal1998ApJ...504..915B} and 0.3 pc, respectively. 
 The latter was estimated considering an angular diameter of $\sim 1'$, and a distance 
 of $\sim 1$ kpc, which is the average value corresponding to the PN scenario.
 One of the characteristics usually associated with proto-PNe and PNe is the presence of H$_2$
 emission in the near infrared. 
 \citet{garcia-hernandez2002} and \citet{kelly2005ApJ...629.1040K} 
 found that almost all the B-type stars in their proto-PNe samples
 showed emission from the $2.122$ $\mu$m H$_2$ line, with the exception
 of one object in each sample.
 Furthermore, \citet{kelly2005ApJ...629.1040K} found that all the B
 stars that showed H$_2$ emission also presented Br$\gamma$ emission,
 although in most cases this line was weaker than the $2.122$ $\mu$m line.
 MWC\,137 exhibits the \,Br$\gamma$ line in strong emission, but no H$_2$ lines have been
 detected \citep{oksala2013}, 
 which might suggest that the star is not in a proto-PN phase.  
 However, \citet{garcia-hernandez2002} and \citet{kelly2005ApJ...629.1040K} proposed several
 explanations for the lack of detection of H$_2$ emission in the B-type stars included in
 their samples (e.g., misplacement of the slit, weak emission due to photodissociation),
 so this scenario cannot be ruled out yet. 
  
 Near-infrared CO band emission has been detected in the spectra of post-AGB stars 
 \citep[e.g.][]{hrivnak1994,garcia-hernandez2002,oudmaijer1995,gledhill2011} 
 and proto-PNe \citep{hora1999}, indicating that this molecule can be present in the 
 circumstellar environment of these objects. 
 Compact molecular disks have been detected around post-AGB stars and proto-PNe, 
 and although the material is usually found to be in expansion, rotation has been 
 proposed in several cases, and confirmed at least in one object, the proto-PN HD 44179, 
 known as the Red Rectangle nebula \citep{bujarrabal2013A&A...552A.116B}. 
 The low luminosity position of MWC\,137 in the HR diagram and the derived value of the 
 carbon isotopic ratio are compatible with a PN stage only for a very low initial rotation rate. 
 If the CO band emission observed in the spectrum of MWC\,137 comes from a rotating ring 
 with a velocity, projected to the line of sight, of $84 \pm 2$ km$\,$s$^{-1}$, then the 
 question is how did the disk form if the star was initially rotating very slowly. 
 Although binarity could lead to the formation of a disk, no companion has been detected so far 
 \citep{baines2006MNRAS.367..737B,wheelwright2010MNRAS.401.1199W}. 
 Magnetic fields have also been proposed to cause asymmetries in the ejected material, 
 but to our knowledge no magnetic field measurements exist for MWC\,137 to support 
 or deny this possibility.
 On the other hand, while the shape of the first \textsuperscript{12}CO band head 
 (blue shoulder and red peak) is typically associated with rotation, it can also result from
 an equatorial outflow of constant velocity. Hence the kinematics in the CO band emission 
 in MWC\,137 could be consistent with the expansion seen in most proto-PNe.
 Still the question of how the material was confined to the equatorial plane remains.

\subsubsection{Main-sequence Star}

 If the star is still on the main sequence, the surrounding nebula should be composed 
 of swept-up interstellar material, rather than material ejected by the star. 
 The position in the HR diagram consistent with a main-sequence classification is 
 associated with a distance $d = 1.3$ kpc, and since the total size of the nebula in the 
 H$\mathrm{\alpha}$ images is $\sim 1'$, the radius of the structure is $\sim 0.2 $ pc. 
 From the evolutionary tracks of \citet{ekstrom2012}, we estimate an age of 10 Myr 
 for a 12 M$_{\sun}$ star, and 7 - 8 Myr for a 15 M$_{\sun}$ star.  
 According to recent simulations, a star with an initial mass of 15 M$_{\sun}$ and 
 an age of 7 Myr would have created a bubble with a radius of $\sim 10 $ pc
 \citep{georgy2013A&A...559A..69G}, and a similar result is expected for a slightly 
 lower mass star. The size of the nebula around MWC\,137 is much too small to be 
 consistent with a scenario of swept-up interstellar material.
  
 The CO band emission originates in a narrow ring, which according to the temperature 
 derived from modeling, is detached from the star but relatively close.
 One issue with this scenario is the unlikely possibility of early-type main-sequence 
 stars to form and maintain such high-density disks, as strong winds and ionizing radiation
 make the survival of these structures improbable.
 The evaporation timescale of pre-main-sequence disks around Herbig Be stars is typically 
 less than 1 Myr \citep{alonso-albi2009}. 
 For this reason, disks are usually found around main-sequence stars of later spectral 
 types (A to G). However, \citet{stolte2010ApJ...718..810S} found a significant number 
 of B-type main-sequence stars with dusty disks in the Arches cluster, and three 
 of these stars even show first overtone CO band emission in their $K$-band spectra.
 The cluster has an age of $2.5 \pm 0.5$ Myr \citep{najarro2004ApJ...611L.105N}, which 
 is much longer than the evaporation timescale, but still might be compatible with a 
 pre-main-sequence origin of the disks \citep{olczak2012ApJ...756..123O}. 
 However, given the age range of MWC\,137, such an old B-type main-sequence star
 should definitely have lost its primordial disk. 
 
 Further, the presence of chemically enriched material in the ring around MWC 137 
 is contrary to what is expected from disks around main-sequence stars.
 The latter should consist of unprocessed matter from the star formation process. 
 Consequently, the material forming the disk around MWC\,137 must be of stellar origin.
 The observed carbon enrichment is consistent with a main-sequence stage of a star 
 with a relatively high initial rotation rate. 
 One of the possibilities for disk formation is related to critical rotation, 
 such as for classical Be stars 
 \citep[e.g.][]{krticka2011A&A...527A..84K,kurfust2012ASPC..464..223K,granada2013A&A...553A..25G}. 
 However, the densities in Be star disks are usually not high enough for efficient molecule 
 and dust condensation, excluding MWC\,137 as a classical Be star.
 Therefore, the only mechanism that can result in a high-density chemically enriched disk 
 around an early-B main-sequence star is interaction with a close binary companion, 
 or even full binary merger.
 Merging of close massive binaries during the main-sequence evolution is predicted 
 by binary evolution models \citep{sana2012Sci...337..444S}.
 A merger scenario could explain the lack of detection of a companion star.

\subsubsection{Supergiant}

 Considering the lower limit of 6 kpc for the distance estimated by \citet{esteban1998}, 
 and the corresponding lower limit of $\log (L/L_{\sun}) = 5.4$ for the luminosity, 
 the value of the $ \mathrm{ \textsuperscript{12}C / \textsuperscript{13}C }$ ratio derived 
 from the spectral modeling is consistent with a rotating star that has just left the 
 main-sequence and is now entering the supergiant phase. 
 There are two well-known early B-type supergiants in the Galaxy that can be compared 
 to MWC\,137: Sher 25 and SBW1.
 Their optical images show bipolar lobes and equatorial rings with estimated radii 
 of $\sim 0.2$ pc \citep{brandner1997ApJ...475L..45B,smith2007AJ....134..846S}. 
 The abundance determinations indicate that these stars did not go through a red 
 supergiant (RSG) phase yet, and that the ring material has most likely been ejected 
 during the blue supergiant (BSG) phase \citep{hendry2008MNRAS.388.1127H,smith2007AJ....134..846S}.
 \citet{hendry2008MNRAS.388.1127H} analyze the radial velocity variations of Sher 25 
 and conclude that it is unlikely that this star has an unseen companion. 
 For SBW1 no evidence of binarity has been reported yet.  
 Instead, it was suggested that the formation of the rings might be connected 
 to rapid rotation of the central stars.
 The H$\alpha$ image of MWC\,137 shows a structure that could also be interpreted 
 as a bipolar nebula \citep[see Fig.\,2 from][]{marston2008}. 
 If the projection of the lobes on the plane of the sky have an apparent
 size of $\sim 1'$ \citep[see Fig.\,3 from][]{esteban1998}, for the distance 
 of 6 kpc the linear radius would be $\sim 1$ pc. 
 
 While on large scales the three objects are comparable, on smaller scales 
 more similarities can be found between SBW1 and MWC\,137.
 The CO emission in MWC\,137 must come from a ring close to the star.
 Nothing is known about molecular band emission from the other BSG, however, 
 \citet{smith2013MNRAS.429.1324S} found a narrow and dense torus of warm dust around SBW1.
 This may suggest that this star could be similar to MWC\,137, but in a more evolved stage, 
 in which the initially hot molecular material has expanded, cooled and condensed into dust. 
 This would be compatible with the lack of a dusty disk or ring structure around MWC\,137.
 An additional similarity between these two stars is the fact that in both cases the
 [O\,{\sc iii}] emission from the nebula is very weak \citep{smith2013MNRAS.429.1324S}.

 Since MWC\,137 exhibits typical characteristics associated with the B[e] phenomenon,
 \citet{esteban1998} propose that the star is a B[e] supergiant (B[e]SG). 
 These stars are known to have dense molecular disks or rings, and first overtone 
 CO band emission has been observed in many of them. 
 Detailed modeling of this emission indicates the existence of 
  detached rotating molecular rings with temperatures and column densities comparable 
 to the values derived for MWC\,137 around several B[e] supergiants
 \citep[e.g.][]{kraus2010A&A...517A..30K,liermann2010,cidale2012,oksala2013,kraus2014}.
 Furthermore, an enhancement of the \textsuperscript{13}C abundance has been found 
 in these objects, and the values of the carbon isotopic ratio derived by modeling 
 their $K$-band spectra are, in most cases, consistent with the stars being in a 
 pre-RSG phase \citep{oksala2013}.
 \citet{smith2007AJ....134..846S} suggest that B[e]SGs could be the progenitors of
 early BSGs with rings, such as SBW1 and Sher 25. 
 Considering the possibility that MWC\,137 is a B[e]SG, its close resemblance to these
 two early B supergiants might support this idea.
 However, according to the position in the HR diagram the enrichment in \textsuperscript{13}C 
 seen in the spectrum is not consistent with a high initial rotation rate, thought to be the 
 mechanism for disk formation in B[e]SGs. Also, as no companion has been detected, the formation
 of the molecular ring and the large nebular structure remain unclear.

 \subsubsection{On the Evolutionary Stage of MWC\,137}
 
 We have presented a brief comparison between MWC\,137 
 and objects in each of the evolutionary phases
 consistent with the carbon isotopic ratio derived from our data.
 As was already mentioned, this star shows some characteristics in common with proto-PNe,
 although the main argument against this classification might be the lack of H$_2$ emission.
 However, if MWC\,137 were indeed a proto-PN, it would be one of the relatively few B-type
 stars found in this transition stage, making it an ideal candidate to help 
 identify the characteristics that define these objects and test the stellar evolution
 theories. Besides, the same unique characteristics that attempt against its classification as
 proto-PN could help understand the processes involved in the transition from the AGB to the PN phase.
 The main-sequence classification might be the least likely, given that no compact disks 
 have been confirmed so far around hot main-sequence stars in the age range estimated for MWC\,137.
 The presence of chemically enriched material in the disk of MWC\,137 together with the 
 peculiar H$\alpha$ nebula that surrounds the star complicates the main-sequence picture even more.
 If MWC\,137 were a main-sequence object, a careful study of the mechanisms that led
 to the origin and characteristics of the circumstellar environment could provide clues
 to the processes that can affect mass loss during this phase of the evolution.
 Given the similarities with other BSGs with bipolar ring nebulae and B[e]SGs stars 
 (e.g., bipolar structure, weak [O\,{\sc iii}] emission, high-density disk,
 enhancement of \textsuperscript{13}C) 
 the supergiant classification seems to be the most plausible. 
 Supporting the supergiant scenario, the profiles of the Pfund
 lines suggest that they probably originate in a wind or outflow.
 The presence of a wind is associated with a mass-loss process that
 could be responsible of the enrichment of the circumstellar medium.
 

\section{CONCLUSIONS}

 In this paper, we have presented high-resolution $K$-band spectra of the Galactic B[e] 
 star MWC\,137, aimed at studying the molecular component of the circumstellar medium.
 The low temperature derived by modeling the CO band head features ($T = 1900\pm 100$\,K) indicates that the 
 emitting region is not located in a disk that extends down to the stellar surface, 
 but rather in a narrow, detached ring. This ring seems to be stable, with a
 rotational velocity, projected to the line of sight, of $84 \pm 2$ km$\,$s$^{-1}$.
 The temperature and column density ($N = (3\pm 1)\times 10^{21}$\,cm$^{-2}$) of the disk
 resemble those typically found around B[e]SGs.

 We also investigate the evolutionary stage of this peculiar object based on the carbon isotopic ratio.
 The amount of \textsuperscript{13}CO relative to \textsuperscript{12}CO 
 in the near-infrared spectrum of MWC\,137 is a 
 definite sign of evolution, and confirms that the star cannot be in a pre-main-sequence phase, 
 thus excluding the Herbig Be classification. 
  Considering the classifications reported in the literature and
 the variation of the $ \mathrm{ \textsuperscript{12}C / \textsuperscript{13}C }$
 ratio along the stellar evolution, we found that the observed isotopic abundance ratio
 derived from our modeling is compatible with a proto-PN, main-sequence or supergiant evolutionary phase.
 Furthermore, based on the comparison of observable characteristics of 
 MWC\,137 with objects in each of these phases,
 we propose that the supergiant scenario seems to be the most plausible. 
 Hence, we suggest that MWC\,137 could be in an extremely short-lived phase
 evolving from a B[e] supergiant to a blue supergiant with a bipolar ring nebula.

\acknowledgements

We thank the referee for the useful comments and suggestions which
helped to improve this manuscript.
This research has made use of NASA's Astrophysics Data System (ADS),
and the SIMBAD database, operated at CDS, Strasbourg, France.
M.L.A., L.C. and M.F.M. acknowledge financial support from the Agencia de Promoci\'on
Cient\'ifica y Tecnol\'ogica (Pr\'estamo BID PICT 2011/0885), CONICET (PIP 0300),
and the Programa de Incentivos (G11/109) of the Universidad Nacional de La Plata, Argentina.
M.K. acknowledges financial support from GA\v{C}R under grant number 14-21373S. 
M.E.O. acknowledges the post-doctoral program of the Czech Academy of Sciences. 
The Astronomical Institute Ond\v{r}ejov is supported by the project RVO:67985815.
Financial support for International Cooperation of the Czech Republic
(M\v{S}MT, 7AMB14AR017) and Argentina (Mincyt-Meys, ARC/13/12 and CONICET/14/003) is acknowledged.

\end{document}